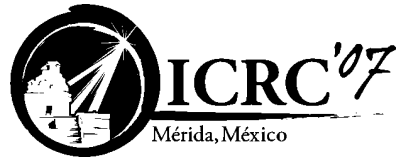

# The ANTARES detector: background sources and effects on detector performance

S. ESCOFFIER[1] ON BEHALF OF THE ANTARES COLLABORATION
[1]Centre de Physique des Particules de Marseille, CNRS/IN2P3
163, av. de Luminy, Case 902, 13288 Marseille Cedex 09, France
escoffier@cppm.in2p3.fr

**Abstract:** The ANTARES Collaboration is deploying a large neutrino detector at a depth of 2475 m in the Mediterranean Sea, 40 km off shore from La Seyne-sur-Mer in South France. The construction of this 12-line detector with 75 phototubes per line will be completed early 2008. Data taking has begun since April 2005 with an instrumentation line also equipped with optical modules. The first 5 detector lines are operational since January 2007. The telescope is aimed to observe high energy cosmic neutrinos through the detection of the Cerenkov light produced by up-going induced muons. Background sources are due to atmospheric neutrinos as well as misreconstructed atmospheric muons. Additional backgrounds inherent to the sea water environment come from $^{40}$K decay and marine organisms' luminescence. While the contribution of the former is expected to be constant at a level of about 45 kHz, the bioluminescence has shown large time variations, with periods of very high activity, up to several hundred kHz. Description of these background sources will be reported, and effects on detector performance will be described. Methods recently developed to improve the detection efficiency in high background periods will be described, together with some of the results obtained.

## Introduction

The goal of the ANTARES neutrino telescope is the observation of high energy cosmic neutrinos from astrophysical sources or originating from dark matter decays. Neutrinos interacting in the vicinity of the detector produce a charged muon whose direction is reconstructed from the arrival times of the Cerenkov photons detected in a large array of photomultipliers [1].

Atmospheric neutrinos and muons are the principle backgrounds for reconstructed events. The former can be rejected using the knowledge of the source direction, and possibly using timing information for transient sources. Concerning the latter, zenith angle cuts should discriminate upward going muons induced by neutrinos, even though some downward going atmospheric muons can be erroneously reconstructed as upgoing tracks [2]. In the case of a diffuse flux search, backgrounds can be suppressed thanks to the harder spectrum expected from cosmic accelerators at high energy.

To these backgrounds is added the intrinsic noise of undersea neutrino telescopes, which comprises the contribution of the $^{40}$K decays and the marine bioluminescence. Since the underwater connection of the first ANTARES lines in March 2006, the impact of this background on the detector performances has been studied.

## Background sources

Counting rates from atmospheric neutrinos and muons are expected to be on the order of a few Hz in the ANTARES detector, which means that bioluminescence and $^{40}$K are locally the dominant background for a photomultiplier. Figure 1 shows the counting rate averaged over 1 minute for all the Optical Modules (OM) in all the five lines and figure 2 shows a typical singles rate counting sequence for Line 4.

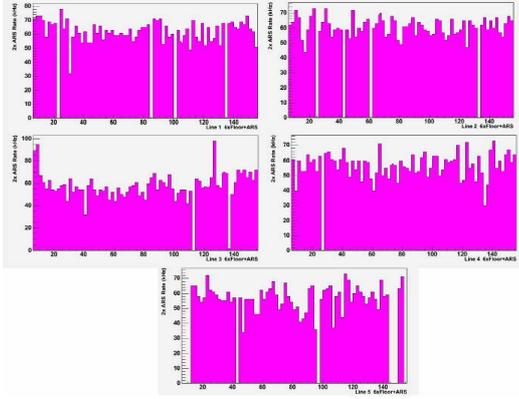

Figure 1: Distribution of counting rates on 22 February 2007 in each of the five lines, giving the counting rate per optical module in kHz for each of the 75 OMs in each line. The lines 3-5 are lacking calibration and so have variations in rates due to different thresholds. The holes usually correspond to dead channels but in this data, in line 5, two storeys were absent from the readout and also appear as holes.

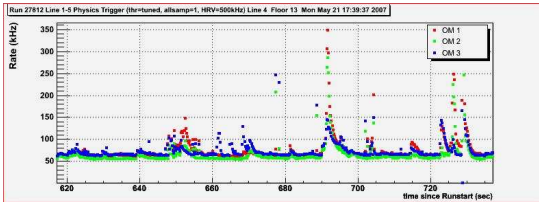

Figure 2: Online counting rates on the three Optical Modules of the floor 13 of line 4 during two minutes on 21st of May 2007.

The continuous part of the counting rates includes the contribution of the $^{40}$K as well as the light emission from bacteria, whereas the light bursts are due to other animal species.

**Potassium-40 decay**

Among salts from sea water, $^{40}$K is the most natural radioactive element. With a half-time of about 1.277 Gyr, this isotope decays in two dominant modes, whom both usually result in Cerenkov radiation:

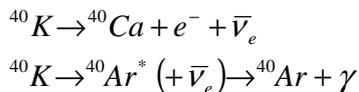

In sea water, the $^{40}$K activity is derived from the measured salinity, which is monitored on the MILOM instrumentation line [3] and is quite stable, within 0.1%. Since the physics involved is well known, its contribution is easily simulated by Monte Carlo and is predicted to be of the order of R=34±7 kHz, including the OM dark counts.

Moreover, $^{40}$K can be a useful tool to monitor OM efficiency, since the secondary electrons have an average path length of only a few millimeters, with many Cerenkov photons produced in a single decay. Cerenkov light can then be seen in coincidence mode on two Optical Modules of the same storey. Figure 3 shows the distribution of time delay between hits on OM0 and OM2 of one storey. This distribution consists of a flat pedestal and a small peak. The pedestal is due to random background coincidences from both $^{40}$K and bioluminescence, whereas the peak is explained as genuine coincidences from $^{40}$K, yielding a rate of 14 Hz. This result is in good agreement with simulation, which predicts a coincidence rate of 13±4 Hz, and implies a $^{40}$K constant single rate in the order of 40 kHz.

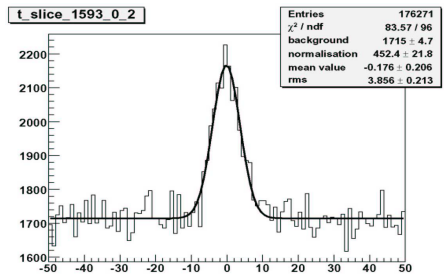

Figure 3: Distribution of time delay between hits recorded by OM0 and OM2 for one storey.

**Bioluminescence**

The baseline rate, the continuous part of the PMT counting rates, computed for each 15 minutes data set, can be seen as a low-pass filter. Contribution of $^{40}$K being constant over time, variations of baseline are assumed to be due to bioluminescence from bacteria, since bacteria emit light continuously. Moreover, water samples obtained during sea campaigns and carried out by biologists from CNRS/INSU showed that bioluminescent bacteria are present at 2475 m of depth on the ANTARES site.

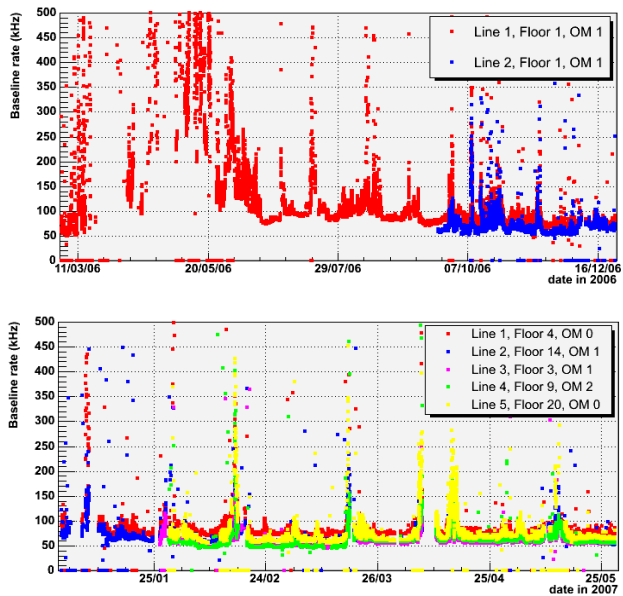

Figure 4: Baseline rates over 9 months in 2006 (top) and 6 months in 2007 (bottom). Sea connections occurred in March and September 2006 for lines 1 and 2 respectively, and at the end of January 2007 for lines 3, 4 and 5.

Baseline rate is shown on figure 4 for 15 months of data taking. Periods with high baseline counting rates, presumably due to high bioluminescence activity, have been observed during the spring of 2006.

Other animal species are also involved in bioluminescence processes, characterized by light bursts emitted spontaneously or by interaction with the ANTARES detector. Figure 5 shows the burst time fraction as a function of the water current.

## Effects on detector performance and trigger strategy

In order to minimize the impact of bioluminescence on the electronic acquisition [4], a cut off is applied as soon as counting rates exceed 500 kHz. The percentage of missing frames is shown on figure 6 as a function of baseline rates. In 2007, it never exceeds 6 % of acquisition time.

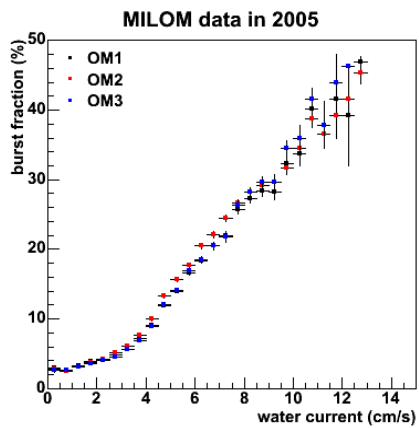

Figure 5: Fraction of time where counting rates are higher than 20% over the baseline as a function of the water current.

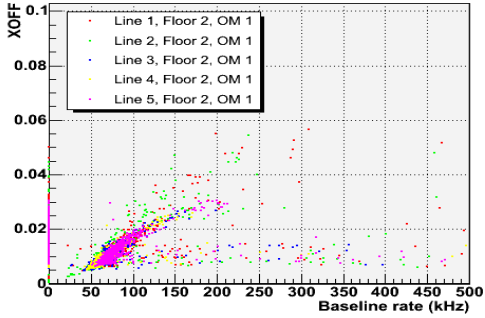

Figure 6: Percentage of missing frames (labeled Xoff) as a function of the baseline rate, during 6 months of data taking in 2007.

Study of the impact of bioluminescence on detector performance requires detailed simulations for each physics channel. Indeed, impact will be different for diffuse sources, which require a high energy cut and no pointing accuracy, for point sources, which needed a high pointing accuracy, for transient sources, for which background is reduced due to time constraint, or for dark matter search, in the low-energy regime. By way of example, the performance of the ANTARES detector is shown on figure 7 in the sensitive range of low energies.

Complementary studies are underway to optimize trigger and reconstruction efficiency at high bioluminescence levels.

## Conclusions

The ANTARES 5-line detector is operational since January 2007. Studies are now in progress to evaluate the impact of bioluminescence on track reconstruction efficiency and on detector performance. Methods are being developed, in particular to improve the data acquisition and trigger conditions at high background counting rates.

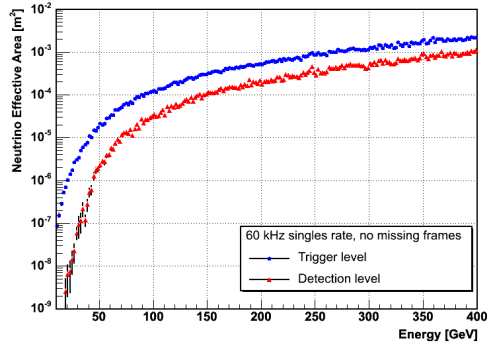

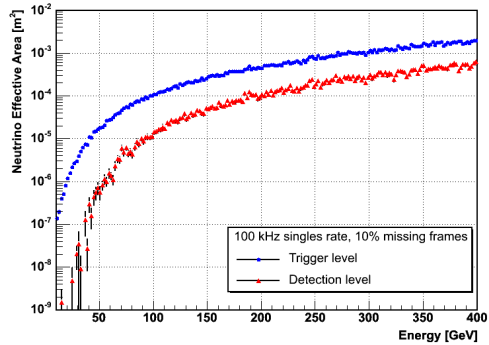

Figure 7: Neutrino effective area as a function of neutrino energy, at the trigger level (in blue) and after track reconstruction (in red), for a nominal 60 kHz singles rate and with no missing frame (top panel), and for a 100 kHz singles rate and 10% of missing frames (bottom panel).